\documentclass[aps, prx, twocolumn,show pacs,show keys,superscript address,superscript reference,floatfix]{revtex4-1}
\usepackage[colorlinks=true,citecolor=blue,linkcolor=blue,urlcolor=blue]{hyperref}
\usepackage[sort&compress]{natbib}
\usepackage{graphicx,times}
\usepackage{color}
\usepackage{amssymb}

\usepackage{subeqnarray}
\usepackage{float}
\usepackage{bbm}
\usepackage{bm}
\usepackage{diagbox}
\usepackage{makecell}
\usepackage{amsmath}

\DeclareMathAlphabet\mathbfcal{OMS}{cmsy}{b}{n}
\newcommand{\ket}[1]{\ensuremath{\left|{#1}\right\rangle}}
\newcommand{\bra}[1]{\ensuremath{\left\langle{#1}\right|}}

\begin{document}
\title{Entangled Quantum Memristors}
\author{Shubham Kumar}
\affiliation{International Center of Quantum Artificial Intelligence for Science and Technology (QuArtist) \\
and Physics Department, Shanghai University, 200444 Shanghai, China} 

\author{Francisco~A.~C\'ardenas-L\'opez}
\affiliation{International Center of Quantum Artificial Intelligence for Science and Technology (QuArtist) \\
and Physics Department, Shanghai University, 200444 Shanghai, China} 

\author{Narendra.~N.~Hegade}
\affiliation{International Center of Quantum Artificial Intelligence for Science and Technology (QuArtist) \\
and Physics Department, Shanghai University, 200444 Shanghai, China} 

\author{Xi~Chen}
\affiliation{Department of Physical Chemistry, University of the Basque Country UPV/EHU, Apartado 644, E-48080 Bilbao, Spain}

\author{Francisco~Albarr\'an-Arriagada}
\email[Francisco~Albarr\'an-Arriagada]{\qquad pancho.albarran@gmail.com}
\affiliation{International Center of Quantum Artificial Intelligence for Science and Technology (QuArtist) \\
and Physics Department, Shanghai University, 200444 Shanghai, China} 

\author{Enrique~Solano}
\email[Enrique Solano]{\qquad enr.solano@gmail.com}
\affiliation{International Center of Quantum Artificial Intelligence for Science and Technology (QuArtist) \\
and Physics Department, Shanghai University, 200444 Shanghai, China} 
\affiliation{Department of Physical Chemistry, University of the Basque Country UPV/EHU, Apartado 644, E-48080 Bilbao, Spain}
\affiliation{IKERBASQUE, Basque Foundation for Science, Plaza Euskadi, 5, 48009 Bilbao, Spain}
\affiliation{Kipu Quantum, Kurwenalstrasse 1, 80804 Munich, Germany}

\author{Gabriel~Alvarado~Barrios}
\email[Gabriel~Alvarado~Barrios]{\qquad phys.gabriel@gmail.com}
\affiliation{International Center of Quantum Artificial Intelligence for Science and Technology (QuArtist) \\
and Physics Department, Shanghai University, 200444 Shanghai, China} 

\begin{abstract}
We propose the interaction of two quantum memristors via capacitive and inductive coupling in feasible superconducting circuit architectures. In this composed system the input gets correlated in time, which changes the dynamic response of each quantum memristor in terms of its pinched hysteresis curve and their nontrivial entanglement. In this sense, the concurrence and memristive dynamics follow an inverse behavior, showing maximal values of entanglement when the hysteresis curve is minimal and vice versa. Moreover, the direction followed in time by the hysteresis curve is reversed whenever the quantum memristor entanglement is maximal. The study of composed quantum memristors paves the way for developing neuromorphic quantum computers and native quantum neural networks, on the path towards quantum advantage with current NISQ technologies.
\end{abstract}

\maketitle

\section{Introduction}
\label{sec.1}
The memristor was proposed as the fourth basic circuit element with a flux-charge relation~\cite{IEEE.1971,IEEE.1976}. The first experimental memristor was claimed by HP~labs~\cite{Nat.2008}, which turned into a growing interest in the field. A key feature is its nonlinear current/voltage response exhibiting memory effects, which is used in the area of neuromorphic computing~\cite{Schuman2017}. The goal is to develop a computational paradigm based on neuron-like devices to mimic the human brain capabilities, allowing to simulate from learning processes~\cite{Song.2019, Lin.2019, Gao.2020,Xia.2020} and artificial neural communication \cite{Laughlin2003}, to simulate biological processes such as brain synapses~\cite{Adv.2020,Nat.2020}.

The development of neuromorphic computing has focused into the design of single memristive systems and structures as the cross-bar arrays, where memristors work as switches according to its resistance value~\cite{Borghetti2010}. These structures have been used in dense arrays of memristors \cite{Nat.2015, NatComm.2018} for realizing brain-inspired devices~\cite{Nat.2015,Adv.2020, Davies.2019}. Nevertheless, little attention has been paid into studying the direct coupling between these system, and only few proposals about parallel/series, and wireless connections have been considered~\cite{Budhathoki2013,Luo2017}. Therefore, characterizing the dynamics of coupled memristors will advance neuromorphic computing.

The implementation of neural networks and brain-inspired quantum algorithm on quantum hardware gives rise to the field of neuromorphic quantum computing (NQC), where quantum features such as superposition and entanglement lead to calculation speedup~\cite{Markovica2020}. There exists two NQC approaches, the first one aims at implementing neural networks on quantum processors through parametrized quantum circuits using classical/quantum machine learning techniques~\cite{Cao.2017,Taccinoj.2019,Li.2020,Kristensen.2021}. The second approach, closer to the classical neuromorphic computing, attempts to take advantage of the system dynamics for implementing neuronal and synaptic processes, specifically a quantum memristor. There are several proposals for implementing a quantum memristor in quantum platforms such as photonics~\cite{sanz2018,materials.2020, Spagnolo.2021}, quantum dots~\cite{patent.2017}, and superconducting circuits~\cite{SciRep.2016, PRapp.2016, PRapp.2018,Phys.Rev.Applied.2014,Sci.Rep.2016}, as well as memristor-inspired experiments for quantum advantage~\cite{Zhou2020arXiv}. In the latter, the memristive behavior appears due to the non-linearity of the Josephson junctions (JJs), which contains a current term depending on its superconducting phase~\cite{Joseph.1962}. It is possible to increase this current contribution by coupling two Josephson junctions on a closed-loop, forming a device known as a conductance asymmetric SQUID~\cite {Phys.Rev.Applied.2014,Sci.Rep.2016}. These works have focused on developing a single quantum memristor, and not much attention has been paid to develop and characterize the systems formed by coupled quantum memristors. 

In this work, we show how to entangle two quantum memristors (QMs) via inductive and capacitive couplings within current state of the art superconducting circuit technologies. We observe that the input signals of the memristors get correlated in time, which induces the shrinking and expansion of the hysteresis curves. This behavior is closely related to the rise and decay of entanglement in the coupled system. We find that the entanglement and memristive dynamics follow an inverse behavior, where maximal values of entanglement happen when the form factor of the hysteresis curve is minimal and vice versa. In addition, we find that the coupling causes the direction of the hysteresis curve to be reversed as the relative phase between current and voltage changes in time. Interestingly, this change in direction of the hysteresis curve happens whenever the entanglement between the memristors is maximal. 

The organization of this article is as follows. In Sec. \ref{sec.2} we review the mathematical model of a classical memristor and its generalization to memristive systems. In Sec. \ref{sec.3}, we describe the composite system of quantum memristors. The main part of the manuscript is devoted to the the analysis of the capacitive interaction between the quantum memristors. In Sec. \ref{sec.4}, we describe the system dynamics via coupled differential equations obtained by solving the master equation of the coupled open quantum systems. In Sec. \ref{sec.5}, we show the dynamical reponses of each quantum memristor in the coupled system. In Sec. \ref{sec.6}, we characterize the perfomance of each quantum memristor using a quantity called as the \textit{Form factor}. In Sec. \ref{sec.6}, we show the entanglement and memristive dynamics of the coupled system and investigate the relationship between them. We devote the Sec. \ref{sec.6} to derive and explain the main conclusions of our work. In the Appendices, we provide the technical details on the derivation of the Hamiltonian (Appendix \ref{A}) and the memristive equations (Appendix \ref{B}). Furthermore we have included the results of the inductive interaction (Appendix \ref{C}) and the interaction caused by using an inductor and a capacitor simultaneously (Appendix \ref{D}).
\\
\section{Classical Memristor}
\label{sec.2}
The memristor is a two terminal device with resistive memory effects, in which the resistance depends on the history of the voltage passed through the device. An ideal memristor is fully described by its state-dependent Ohm's law~\cite{IEEE.1971}. Accordingly, based on the intrinsic variables of the device, the memristor can be classified as charge-controlled or flux-controlled with the constitutive relations 
\begin{subequations}
\begin{eqnarray}
V(t) &=& M(q)~I(t) \quad\text{with}\quad M(q)=\frac{d\phi}{dq}\, , \\
 \label{curr}
I(t) &=& G(\phi)~V(t) \quad\text{with}\quad G(\phi)=\frac{dq}{d\phi} \, .
\end{eqnarray}
\end{subequations}
Here, $I(t)$, $V(t)$, $q(t)$, and $\phi(t)$ stand for the current, voltage, charge, and flux on the device. Furthermore, $M(q)$ and $G(\phi)$ are termed as memristance and memductance, with units of electrical resistance and electrical conductance, respectively. The generalization of the device is termed as memristive systems~\cite{IEEE.1976}. The main difference between them corresponds to the explicit dependence of the voltage (current) on the memductance (conductance). The dynamics of the memristive system is given by the state dependent Ohm's law, together with the evolution of the state variable provided by
\begin{subequations}
\begin{eqnarray}
\label{memristive1}
V(t) = M(x,I,t)~I(t), \quad f(x,I,t)=\frac{dx}{dt},\\
\label{memristive2}
I(t) = G(x,V,t)V(t) \quad g(x,V,t)=\frac{dx}{dt}.
\end{eqnarray} 
\end{subequations}
Here, $M(x,I,t)$ and $G(x,V,t)$ are the memristance and memductance, which are now functions of the internal variable $x$ ruled by $ f(x,I,t)$ and $ g(x,V,t)$, respectively.

\begin{figure}[!th]
\centering
\includegraphics[width=1\linewidth]{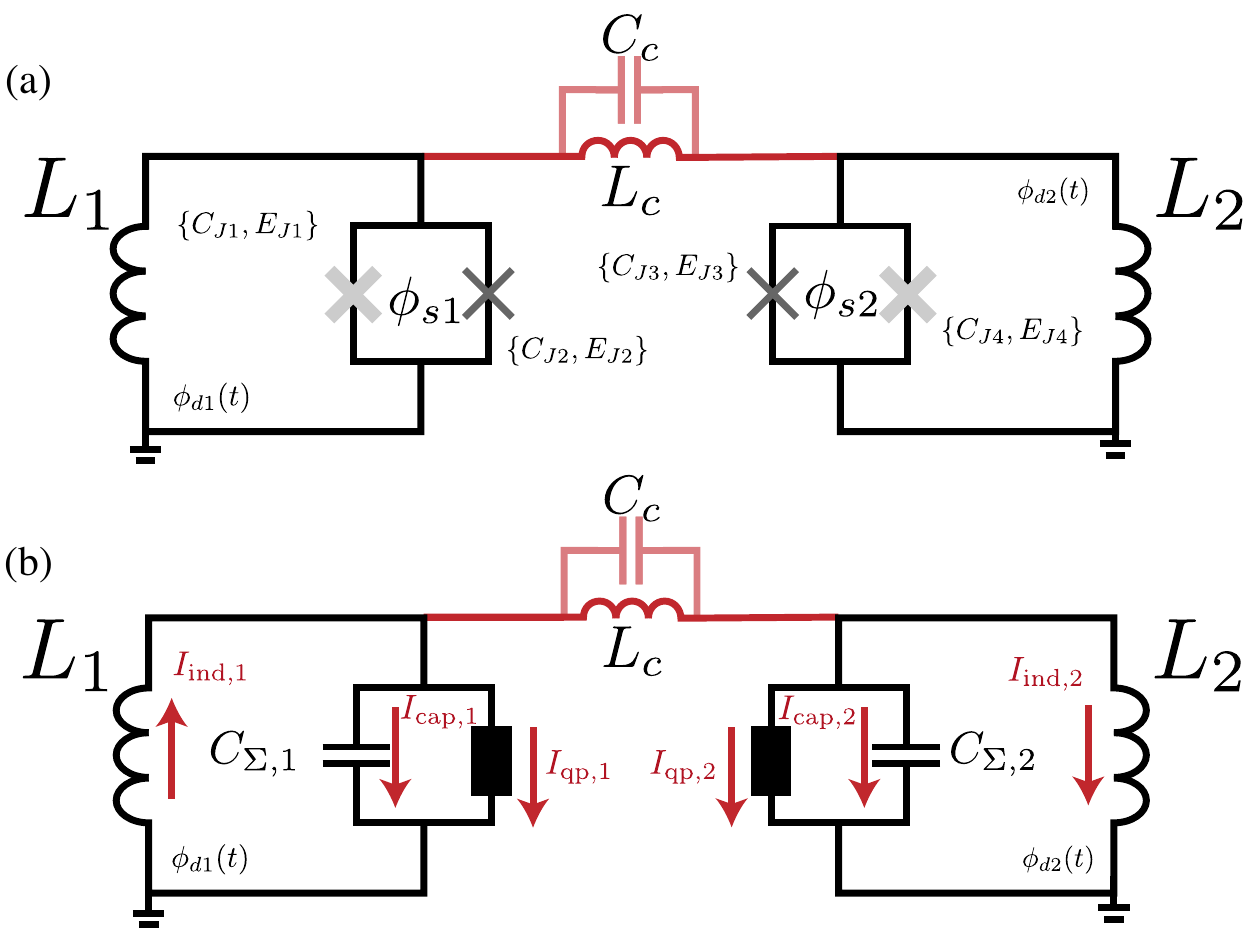}
\caption{Schematic illustration of two coupled superconducting quantum memristors, capacitively and inductively, respectively. (a) Each QM consists of a CA-SQUID connected to an inductance $L_{\ell}$ in a loop threaded by a signal $\phi_{d\ell}(t)$. We describe the CA-SQUID as a smaller loop interrupted by two different junctions and threaded by a static flux $\phi_{s}$, which cancels the critical current. (b) Effective RLC circuit describing the system, where the resistance is modelled as a time-dependent admittance. We depict the current with red arrows.}
\label{Fig:fig1}
\end{figure}

\section{The model}
\label{sec.3}
Let us study the circuit depicted in Fig.~\ref{Fig:fig1}a, consisting of a pair of quantum memristors (QMs) coupled with either a capacitor or an inductor. We form each QM with a conductance asymmetric SQUID (CA-SQUID) connected in parallel to an inductor $L_{\ell}$ forming a closed-loop threaded by an external signal $\phi_{d\ell}(t)$. The CA-SQUID consists of a small loop embedded by two different JJs threaded by a static DC magnetic flux. Each Josephson junction forming the CA-SQUID has two current contributions. The first is a non-dissipative current due the tunnelling of Cooper-pairs proportional to the critical current $I_{C\ell}=2eE_{J}/\hbar$~\cite{Golubov2004RMP}. The second is a dissipative contribution due to the tunneling of quasiparticles proportional to the conductance~\cite{Catelani2011prl,Catelani2011prb}. The memristive behavior arises when the latter current dominates over the former. This is achieved when the junctions satisfy $G_{N1}/G_{N2}=\Delta_1/\Delta_2$~\cite{PRL1963,Phys.Rev.Applied.2014} ($G_{Ni}$ is the junction conductance, $\Delta_i$ is the energy bandgap) and the loop is threaded with a flux $\phi_{s\ell}=\pi$. In the limit $\Delta_1>>\Delta_2$, the first junction acts as a shunt. Then, we model the CA-SQUID as a capacitor coupled to a time-dependent admittance~\cite{Sci.Rep.2016}. Therefore, the whole system corresponds to a pair of coupled RLC resonators with time-dependent resistance (see Fig.~\ref{Fig:fig1}b) described by the Hamiltonian (see Appendix~\ref{A} for derivation)
\begin{eqnarray}
\label{H_2_mem}
\hat{\mathcal{H}} = \sum_{\ell=1,2}E_{C_{\ell,\ell}}\hat{n}^{2}_{\ell}+\frac{E_{L_{\ell,\ell}}}{2}\hat{\phi}_{\ell}^{2}+E_{C_{1,2}}\hat{n}_{1}\hat{n}_{2} - E_{L_{1,2}}\hat{\phi}_{1}\hat{\phi} _{2} \, . \nonumber \\
\end{eqnarray}
Here, $\hat{n}_{\ell}$ and $\hat{\phi}_{\ell} $ are the dimensionless number and phase operators respectively, $E_{C}=2e^2\hat{C}^{-1}$ is the charge energy matrix and $\hat{C}^{-1}$ is the inverse of the capacitance matrix. Moreover, $E_{L}=\varphi_{0}^{2}\hat{L}^{-1}$ is the inductive energy matrix where $\hat{L}^{-1}$ is the inverse of the inductive matrix, and $\varphi_{0}=\hbar/2e$ is the reduced quantum flux. Furthermore, $n_{\ell}$ and $\phi_{\ell}$ correspond to the dimensionless charge and phase operators of the $\ell$th QM, respectively. The first two terms of Eq. (\ref{H_2_mem}) are the free Hamiltonian of each memristive system, whereas the last two terms are the capacitive and inductive coupling interaction, respectively. 

For simplicity we redefine the operators in terms of creation and annihilation operators,
\begin{eqnarray}
\hat{n}_{\ell} &=& \frac{i}{4g_{\ell}}(\hat{a}_{\ell}^{\dagger}-\hat{a}_{\ell}) , \nonumber\\
\hat{\phi}_{\ell} &=& 2g_{\ell}(\hat{a}_{\ell}^{\dagger}+\hat{a}_{\ell}) ,
\end{eqnarray}
with $g_{\ell}=(E_{C_{\ell,\ell}}/32E_{L_{\ell,\ell}})^{1/4}$. Then, we obtain
\begin{eqnarray}
\label{Hamiltonian}
\hat{\mathcal{H}}=\sum_{\ell=1,2}\hbar\omega_\ell \hat{a}_\ell^{\dagger}\hat{a}_\ell - \sqrt{\omega_{1}\omega_{2}}(\alpha-\beta)(\hat{a}_1^{\dagger}\hat{a}_2 + \hat{a}_1\hat{a}_2^{\dagger}),
\label{Eq05}
\end{eqnarray}
where $\omega_{\ell}=\sqrt{2E_{C_{\ell,\ell}}E_{L_{\ell,\ell}}}/\hbar$ is the frequency of the $\ell$th QM. Moreover, parameter $\alpha=E_{L_{1,2}}/\sqrt{E_{L1}E_{L2}}$ and parameter $\beta=E_{C_{1,2}}/\sqrt{E_{C1}E_{C2}}$ is the inductive and charge energy ratios, respectively. We describe the energy loss of the admittance as a quasiparticle bath at zero temperature whose dynamics follows the time-dependent master equation ($\hbar=1$)
\begin{eqnarray} 
\label{Mst_eq_1}
\dot{\hat{\rho}}(t) &=& i\big[\mathcal{\hat{H}},{\hat{\rho}}\big] + \sum_{\ell=1,2}\frac{\Gamma_{\ell}(t)}{2}\bigg[\mathcal{L}_{\ell}\hat{\rho}\mathcal{L}_{\ell}^{\dag}-\frac{1}{2}\{\mathcal{L}_{\ell}^{\dag}\mathcal{L}_{\ell} ,\hat{\rho}\}\bigg] ,
\label{Eq06}
\end{eqnarray}
here, $\mathcal{\hat{H}}$ is the Hamiltonian shown in Eq.~(\ref{Eq05}), while $\mathcal{L}_{\ell}=\sqrt{\Gamma_{\ell}(t)}\hat{a}_{\ell}$ is the collapse operator describing the quasiparticle tunneling of the $\ell$th QMs at rate $\Gamma_{\ell}(t) = \lvert \bra{0}\sin(\hat{\phi}_{\ell}/2)\ket{1}\lvert^{2}S_{\rm{qp}}(\omega_{\ell})$~\cite{Catelani2011prl,Catelani2011prb}. Where, $S_{\rm{qp}}(\omega_{\ell})\approx\omega_{\ell}$ is the spectral density of the quasiparticle bath~\cite{Catelani2011prl,Catelani2011prb}. Due to the fluxoid quantization rule on the outer loop, it is possible to write $\Gamma_{\ell}(t)$ in terms of the external magnetic flux $\phi_{d\ell}(t)$. This means $\Gamma_{\ell}(t) = g_{\ell}^{2}\omega_{\ell}\exp(-g_{\ell}^{2})[(1+\cos[\phi_{d\ell}(t)])/2]$~\cite{Sci.Rep.2016}, we note that $g_{\ell}$ is proportional to the zero-point fluctuation of the phase operator $\hat{\phi}_{\ell}$. Moreover, $\phi_{d\ell}(t)=\phi_{0,\ell}+A\sin(\omega_{\ell}t)$ is the external magnetic flux threading the outer loop, where we consider a sinusoidal modulation for observing the memristive behavior of the coupled system. The phase dependence in the decay rate is due to the quasiparticle current term \cite{PRB.2011}, which is intrinsic to JJs. This allows to control the quasiparticle dynamics via the external flux and thus have a flux controlled memristive device. Since, we are interested in investigating the memristive behaviour originating from the phase dependent conductance, we omit the decoherence induced by other sources such as inductive and radiative losses. Works on superconducting circuits have studied dominant quasiparticle tunneling \cite{PRB.2011, Martinis.2009, Vool.PRL} and recent works have progressed towards mitigating the radiative decay \cite{Martinis.npj} which justifies our assumption.
\begin{figure}[!t]
\centering
\includegraphics[width=0.8\linewidth]{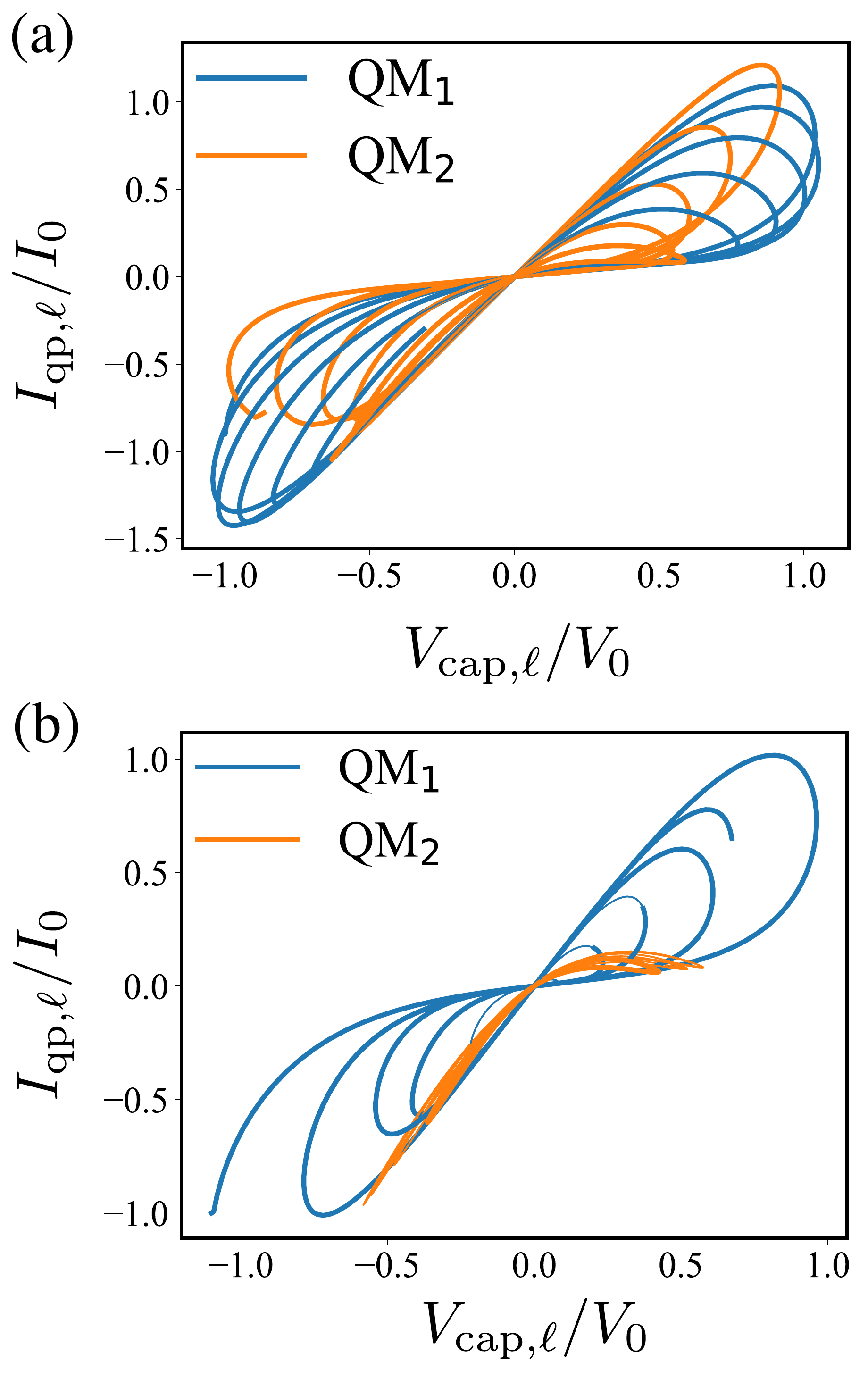}
\caption{Normalized pinched hysteresis curves of identical quantum memristors. QM$_1$ is initialized in the state $\ket{\Psi(\pi/4,\pi/2)}$, whereas QM$_2$ is in the state (a)~$\ket{\Psi(\pi/3,\pi/2)}$ and (b) $\ket{\Psi(\pi/4,0)}$. The system parameters are $C_{\Sigma_{1}}=C_{\Sigma_{2}}=3.6~[\rm{fF}]$, $C_{c}=0.9~[\rm{fF}]$, $L_{1}=L_{2}=6.1~[\rm{\mu H}]$, leading to memristor frequency of $\omega_{1}=\omega_{2}=5.03~[\rm{GHz}]$.}
\label{Fig:fig2}
\end{figure}
\begin{figure}[!t]
\centering
\includegraphics[width=0.8\linewidth]{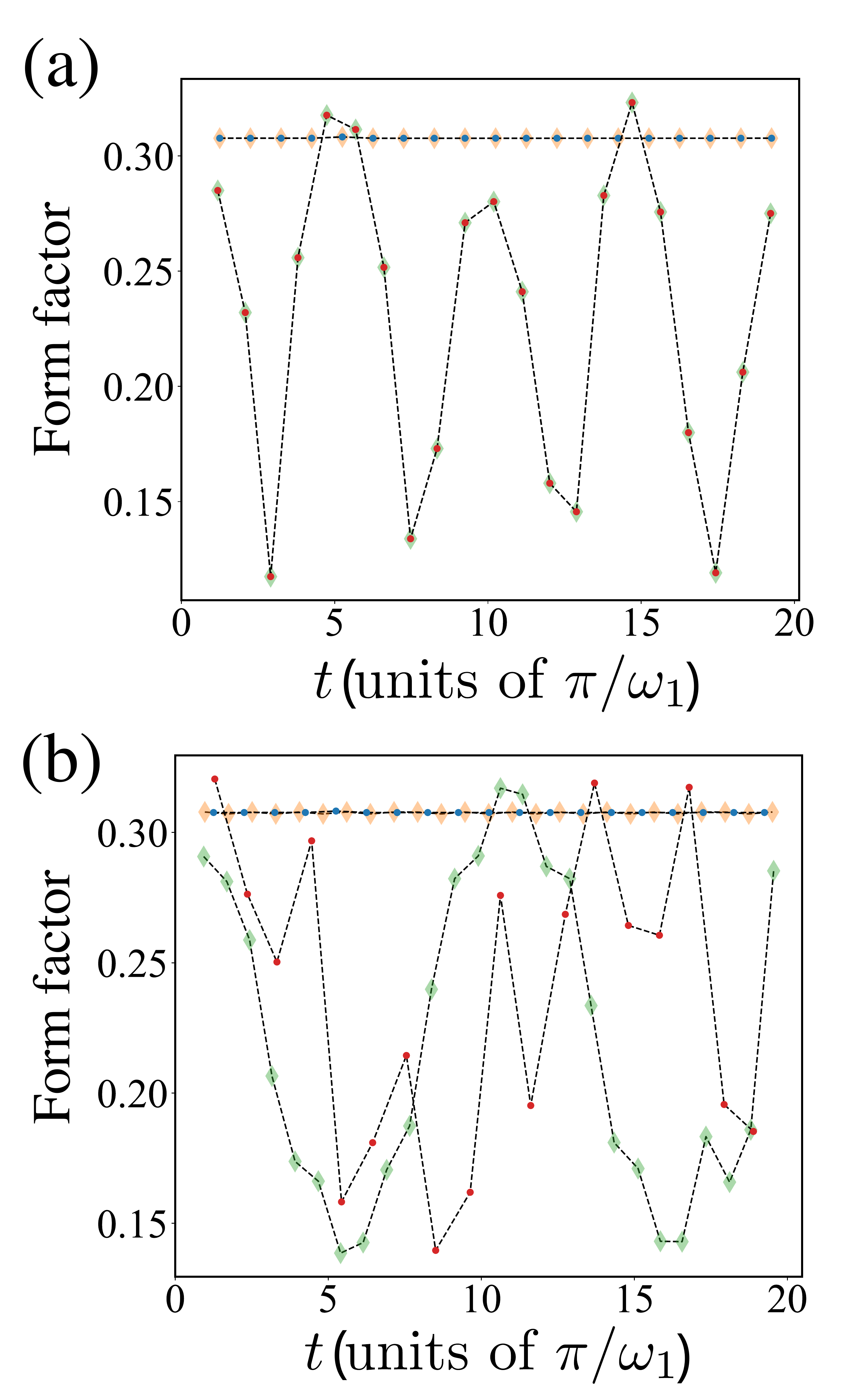}
\caption{Form factor of the hysteresis curves as a function of the period for (a) identical and (b) non-identical QMs. QM$_1$ and QM$_2$ are shown using diamonds and dots, respectively. For identical QMs, the parameters are $\ket{\Psi(\pi/4,\pi/2)}$, $C_{\Sigma_{1}}=C_{\Sigma_{2}}=3.6~[\rm{fF}]$, $L_{1}=L_{2}=6.1~[\rm{\mu H}]$, leading to memristor frequency of $\omega_{1}=\omega_{2}=5.03~[\rm{GHz}]$. For non identical QMs, $\ket{\Psi(\pi/3,\pi/2)}$, $C_{\Sigma_{1}}=3.6~[\rm{fF}]$, $C_{\Sigma_{2}}=2.6~[\rm{fF}]$, $L_{1}=~[\rm{\mu H}]$, $L_{2}=~[\rm{\mu H}]$, leading to a memristor frequency of $\omega_{1}=6~[\rm{GHz}]$, and $\omega_{2}=8~[\rm{GHz}]$.}
\label{Fig:fig3}
\end{figure}

\section{Coupled differential equations}
\label{sec.4}
The current-voltage relation of each QM is obtained through the equation of motion for the operators $\hat{n}_{\ell}$ and $\hat{\phi}_{\ell}$. We can compute them from the master equation of Eq.~(\ref{Eq06}) (see Appendix~\ref{B}), obtaining
\begin{subequations}
\begin{eqnarray}\nonumber
\label{current}
\frac{d}{dt}\langle\hat{n}_\ell\rangle = &&  E_{L_{\ell,\ell}} \langle \hat{\phi}_\ell \rangle -E_{L_{1,2}}(\delta_{1,\ell}\langle \hat{\phi}_2 \rangle-\delta_{2,\ell}\langle \hat{\phi}_1 \rangle)\\
&& - \frac{{\Gamma }_\ell(t)}{2}\langle \hat{n}_\ell \rangle,\\\nonumber
\label{voltage}
\frac{d}{dt}\langle{\hat{\phi}}_{\ell}\rangle = && - 2E_{C_{\ell,\ell}}\langle{\hat{n}}_\ell\rangle -2E_{C_{1,2}}(\delta_{1,\ell}\langle \hat{n}_2 \rangle+\delta_{2,\ell}\langle \hat{n}_1 \rangle) \\
&& - \frac{\Gamma _{\ell}(t)}{2}\langle{\hat{\phi}}_{\ell}\rangle.
\end{eqnarray}
\end{subequations}
We can produce analytical solutions for the averaged quantities $\langle \hat{n}_{\ell}(t)\rangle$ and $\langle \hat{\phi}_{\ell}(t)\rangle$ for identical QMs. For the initial state $\langle \hat{n}_{1}(0)\rangle=\langle \hat{n}_{2}(0)\rangle=n_{0}$ and $\langle \hat{\phi}_{1}(0)\rangle=\langle \hat{\phi}_{2}(0)\rangle=0$, we get
\begin{eqnarray}
\label{current_1}
\langle \hat{n}_\ell\rangle&=&n_{0}\exp\bigg[-\int_{0}^{t} \frac{\Gamma(s)}{2}ds\bigg] \cos( \omega't),\\\nonumber
\label{volt_1}
\langle {\hat{\phi}}_\ell\rangle&=&n_{0}\sqrt{\frac{E_{L_{\ell,\ell}} - E_{L_{1,2}}}{{2(E_{C_{\ell,\ell}} - E_{C_{1,2}}})}}\exp\bigg[-\int_{0}^{t} \frac{\Gamma(s)}{2}ds\bigg]\sin(\omega't).\\
\end{eqnarray}
Here, $\omega'=\sqrt{2(E_{C_{\ell,\ell}} - E_{C_{1,2}})(E_{L_{\ell,\ell}} - E_{L_{1,2}})}=\omega\sqrt{(1-\alpha)(1-\beta)}$ is the effective frequency of the coupled system. From the equation of motion in Eq.~(\ref{current}) and Eq.~(\ref{voltage}), we obtain that the quasiparticle current of each QM is $I_{\rm{qp},\ell}(t)=G_{\ell}(t) V_{\rm{cap},\ell}(t)$. Here, $G_{\ell}=C_{\Sigma\ell}\Gamma(t)/2$ is the conductance of each quantum memristor~\cite{Sci.Rep.2016} and $V_{\rm{cap},\ell}(t)=-2e \langle n_{\ell}(t)\rangle/C_{\Sigma \ell}$ is the voltage across the capacitor in each CA-SQUID. From these expressions, we will study the response of the composite system under variation of the initial input voltage of each QM. Furthermore, we assume that the QMs operate within the two-level approximation satisfying the adiabatic limit~\cite{Phys.Rev.A.2005}. In the following analysis, we will focus on the capacitive coupling. However, we extend the same analysis to the inductive interaction (see Appendix~\ref{C}) and the interaction using an inductor and a capacitor simultaneously (see Appendix \ref{D}).

\begin{figure*}[t]
\centering
\setlength{\belowcaptionskip}{-10pt}
\includegraphics[width=1\linewidth]{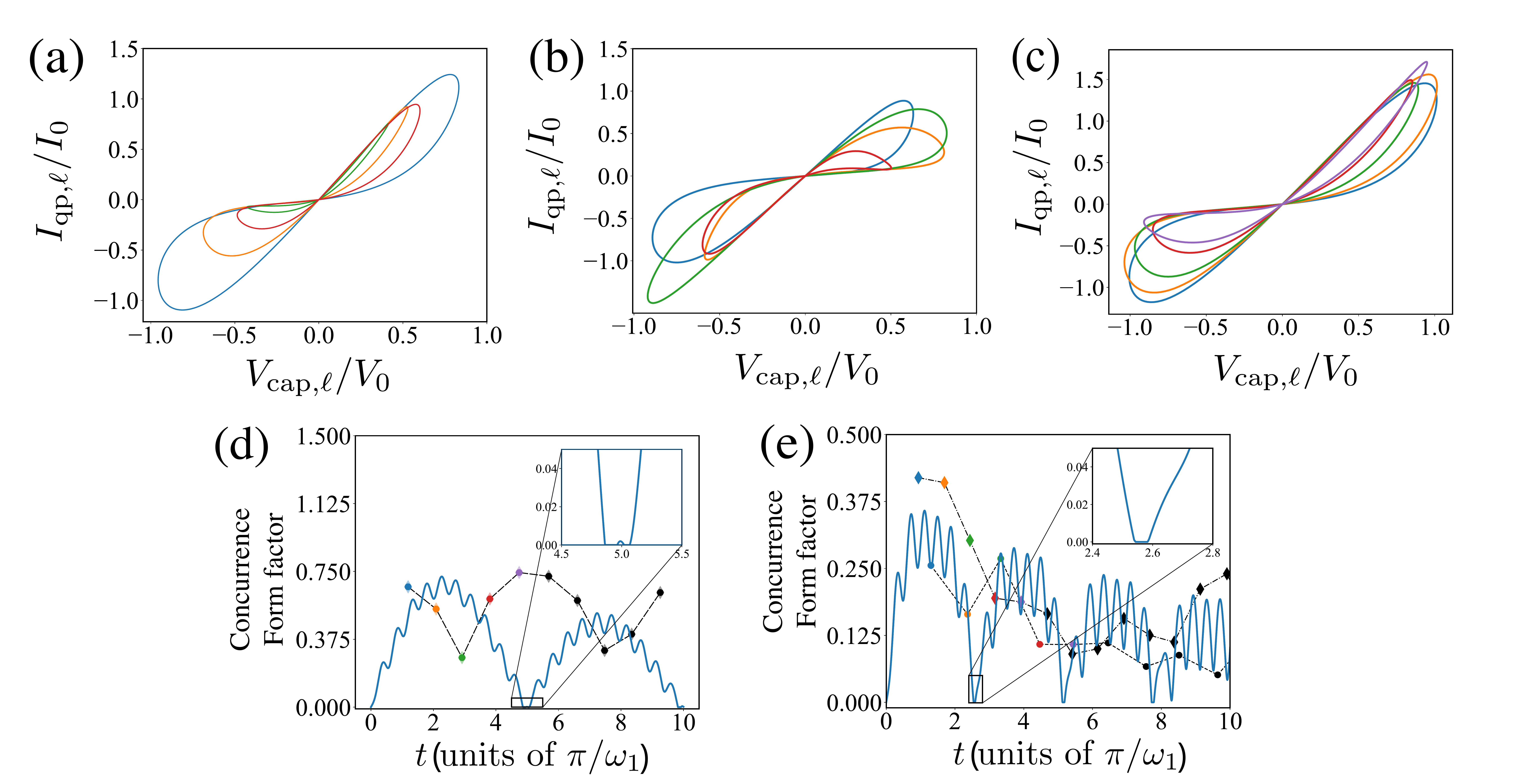}
\caption{Entanglement and memristive dynamics for identical a), (d) and non-identical QMs (b), (c), (e) . We show the memristive dynamics for four oscillations in each case. The form factor (normalized with respect to the concurrence) is plotted along with the concurrence, using the same colors for each oscillation as the hysteresis curves. The form factors for QM$_1$ and QM$_2$ are shown using dots and diamond markers, respectively. Memory revivals can be observed after three oscillations in (d) and after two oscillations for QM$_1$ in (e). For QM$_2$, it can be observed at later times since the dynamics is plotted in the timescale of QM$_1$. The insets show ESD and ESB. We have used the same initial state and system parameters as those in Fig.~\ref{Fig:fig3}.}
\label{Fig:fig4}
\end{figure*}

\section{Dynamic response}
\label{sec.5}
 Let us fix the initial input of the first QM and vary the initial input of the second one. We choose the input through the initial state $\ket{\Psi_{\ell}(\theta_{\ell},\varphi_{\ell})} = \cos(\theta_{\ell}/2)\ket{0}+e^{i\varphi_{\ell}}\sin(\theta_{\ell}/2)\ket{1}$. Depending on the value of $\theta$, it is possible to control the initial voltage/current on the QM, whereas the relative phase $\varphi$ gives us the information whether we initialize the system with an input current ($\varphi=0$) or input voltage ($\varphi=\pi/2$). At the starting point, the composite system is in the product state $\ket{\Psi_1} \otimes \ket{\Psi_2}$. Figure~\ref{Fig:fig2} shows the hysteresis curves for both QMs, where we have initialized the second one with an initial voltage and an initial current, respectively. From Fig.~\ref{Fig:fig2}{a}, as expected, different voltages do not affect the memristive behavior of the device and only the shape of the hysteresis curves change. Opposite to the initial input current case (see Fig.~\ref{Fig:fig2}{b}, we observe that the lobes of the hysteresis curves tends to shrink, with a considerably reduced enclosed area. Thus, the hysteretic behavior of each memristor is very sensitive to initial conditions.

\section{Performance} 
\label{sec.6}
For comparing the performance of each QM in the uncoupled or composite cases, we will focus on their memory effects. It has been shown that the area enclosed by the pinched hysteresis loop is related to the memory effects of the device ~\cite{Radwan2010,Biolek2012,Biolek2014}. Here, we will characterize the memristive features using the form factor given by
\begin{equation}
\mathcal{F}=4\pi\frac{A}{P^2} \, ,
\end{equation}
where, $A$ is the area enclosed by the pinched hysteresis loop and $P$ is its perimeter. By using a dimensionless quantity, that is invariant under scaling, we can compare memory effects and  I/V characteristics despite the decay generated by the dissipative dynamics. We calculate the form factor over consecutive periods where the Lissajous curves reach zero (origin), thus forming a closed loop. Figure~\ref{Fig:fig3} shows the form factor of the hysteresis curves as a function of the period for the uncoupled and coupled QMs. We see that for the uncoupled QMs, the form factor is essentially constant as the area and the square of the perimeter decay at the same rate. This decay appears since the system does not receive additional energy and the input voltage decays due to dissipation, so that the hysteresis follows a constant decay over time. For the coupled QMs, the form factor shows damped oscillations related to the correlation of the inputs. In this case, because the hysteresis curves shrink and expand, it is natural to observe the same pattern in the dynamics of the form factor. For the identical QMs, we see an improvement in the performance of the coupled system. In this manner, the peaks of the form factor overpass the uncoupled case, exhibiting robustness against the loss of its memristive capacities due to dissipation. For the case of non-identical QMs, we can see that there is a trade-off in memristive capacities since there is a phase difference between the oscillations of the form factor of each coupled QMs. This interesting feature manifests since in the timescale of oscillation of any one of them, the decrease and increase in the area of one is compensated by the other. Thus, we observe that the coupling of QMs leads to correlated inputs as the system evolves, inducing periodic decay and revival of the memory.

\section{Entanglement and memristive dynamics}
\label{sec.7}
Following the previous discussions, which hints a relationship between the quantum correlations and the memristive behavior, we investigate the interplay between the entanglement and the memristive dynamics. We compare the memristive dynamics with the evolution of the quantum correlations computed via the concurrence~\cite{PhysRevLett.80.2245,Comp.Phys.2013}  which, for a $2\otimes2$ bipartite state $\rho$ has a closed form
\begin{equation}
\mathcal{C} = \max\{0,\lambda_1-\lambda_2-\lambda_3-\lambda_4\}
\end{equation}
where, $\lambda_i $ are the eigenvalues of the matrix $R =\sqrt{\rho^{1/2}\tilde{\rho}\rho^{1/2}}$ ordered in decreasing order, with $\tilde{\rho} = (\sigma_y \otimes \sigma_y) \rho^* (\sigma_y\otimes \sigma_y)$ where $\sigma_{y}$ is the $y$-component Pauli matrix. $\mathcal{C}$ takes values from zero to one for uncorrelated and maximally correlated states, respectively. Figure~\ref{Fig:fig4} shows the memristive dynamics and the concurrence for the cases of identical ($\omega_1=\omega_2$) and non-identical ($\omega_1\neq\omega_2$) QMs. The selection of system parameters is based on the adiabatic evolution of open quantum systems \cite{Yi.J.Phys} and according to the current state of the art experimental setups for superconducting circuits. The hysteresis plots are shown for the first four oscillations and the form factor is plotted along with the concurrence in the timescale of QM$_1$. Comparing the timescales of the form factor and the concurrence, Fig. \ref{Fig:fig4}d, for the identical QMs we see that the periodic shrink and expansion of the hysteresis curves coincide with the increase and decrease of the entanglement over each period. Accordingly, this shows an inverse relationship between the memristive dynamics and the quantum correlations. Since current and voltage are local observables, an increase in the entanglement leads to a loss of locality, producing hysteresis curves with a smaller area. When the QMs are detuned, Fig. \ref{Fig:fig4}e, the distribution of information exchange between the QMs are uneven due to which the shrink/expansion of the hysteresis and the corresponding rise/decay of the quantum correlations do not happen in the same timescale. This induces a phase shift between the oscillations of the quantum correlations and the memristive behavior.
\begin{figure*}[!t]
\centering
\includegraphics[width=1.0\linewidth]{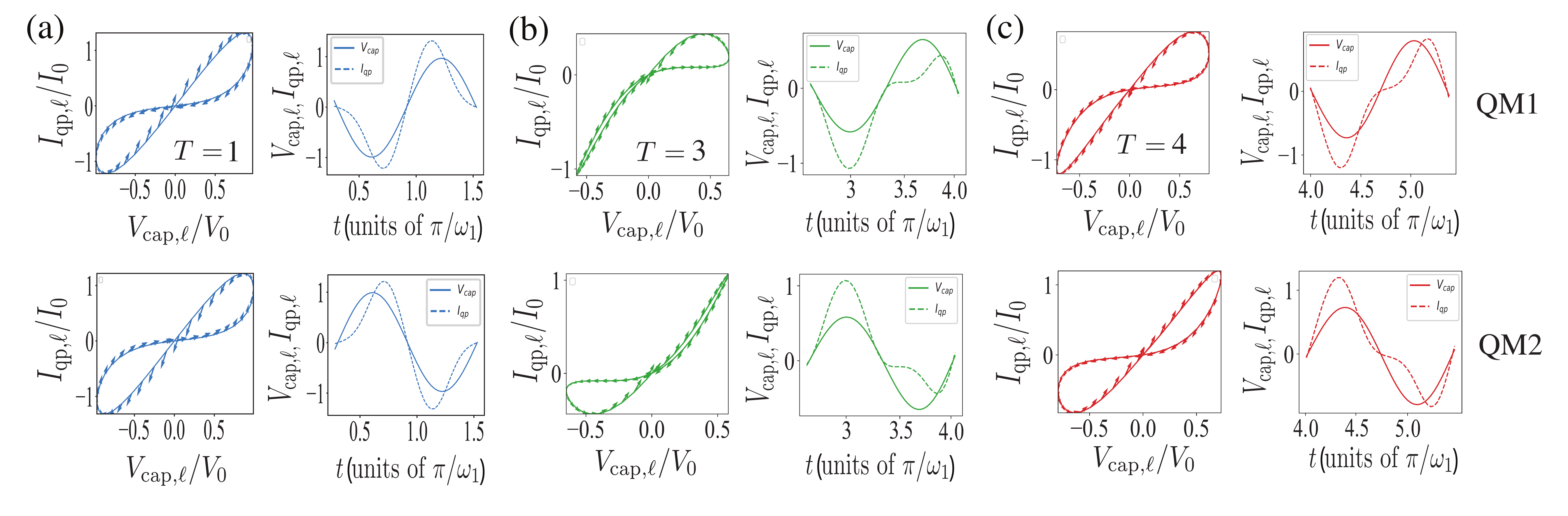}
\setlength{\belowcaptionskip}{-10pt}
\caption{Direction of the hysteresis curve(left) and corresponding time dependence of current and voltage(right) at the (a)~first oscillation, (b)~third oscillation and (c)~fourth oscillation for the case of identical memristors. The first(second) row corresponds to $\textrm{QM}_{1(2)}$. We can see that the relative phase between voltage and current changes in time. In the fourth oscillation (c), this phase has reversed, which changes the direction of the hysteresis curve. We have used the same initial state and system parameters as in the Fig. \ref{Fig:fig3}.}
\label{Fig:fig5}
\end{figure*}
It is interesting to notice that the entanglement dynamics are also related to the direction of the hysteresis curve. For a given memristor, the relative phase between the voltage and current dictates the direction in which the hysteresis curve is drawn in time. In our case, the interaction between memristors makes the relative phase time-dependent. As a consequence, the direction of the hysteresis curve is periodically reversed during the time evolution. The first reversal of the direction of the hysteresis curve is shown in Fig.~\ref{Fig:fig5}. Specifically, in Fig.~\ref{Fig:fig5}a (left), we see the pinched hysteresis loop of both memristors, while the right panels show the corresponding time dependence of voltage and current. Notice that the memristors have opposite direction in their hysteresis curves. We can also see that for each memristor,  as time increases, the relative phase between current and voltage decreases and eventually reverses, which changes the direction of the hysteresis curve, as can be seen in Fig.~\ref{Fig:fig5}c. Interestingly, this change in direction happens when the concurrence between the memristors is maximum, as can be seen by comparing Fig.~\ref{Fig:fig5} with Fig.~\ref{Fig:fig4}a. There, the colors are chosen to correspond to the same oscillations. Then, entanglement reaches its maximum value when the relative phase between current and voltage is minimum.

We also observe that the system exhibits periodic entanglement sudden death (ESD) and sudden birth (ESB)~\cite{PRL.2008}, as shown in the insets of Fig.~\ref{Fig:fig4} for identical and non-identical QMs. The entanglement dynamics in the two coupling schemes differ by timescales and the maximal peak of the entanglement. With detuning, the time taken to reach the death state and the maximal entanglement tends to decrease. ESD/ESB appears due to two identified factors. The first one corresponds to the competition between the two timescales of the system dynamics. Faster oscillations at time $\tau_{1}=2\pi/\omega_{\ell}$ correspond to the updated time of the voltage input of the QM and an enveloping at time $\tau_{2}=2\pi/g$, where $g=\sqrt{\omega_1\omega_2}\alpha(\beta)$, coinciding with the entanglement/disentanglement of the system. The second factor stems from the existence of an effective interaction mediated between subsystem-subsystem, environment-environment, and subsystem-environment. ESB implies the existence of correlations between the system and the environment. The oscillatory behavior of the entanglement and memristive dynamics is a consequence of them, causing a backflow of information. Moreover, the loss of of information due to a decay of quantum correlations shows up in the reservoir, increasing the memory and vice versa. 
Notice that in our system the QMs become entangled because of the nature of their interaction. There are other ways for entangling quantum systems by combination of strong driving and dissipative processes \cite{Ana.PRA}. This is an interesting possibility that could be considered for future works.  

\section{Conclusions}
\label{sec.8}
We have studied the dynamics of two coupled quantum memristors, observing that the capacitive and inductive interactions correlate the input, producing shrink and expansion of the hysteresis loop. As a consequence, periodic decays and revivals of their associated memories emerge. This behavior is inversely related to the quantum correlations generated in the composite quantum system. We find that coupling the memristors changes the relative phase between the current and voltage in time. Therefore, the direction of the hysteresis curve is periodically reversed during the time evolution. We find that the change in direction happens whenever the entanglement reaches a maximal value. In addition, we observe that the composite system exhibits entanglement sudden death and sudden birth due to a combination of the effect of the periodic adiabatic driving and the coupling. 

Compared to classical coupled memristors \cite{Budhathoki2013,Luo2017}, we obtain correlated inputs and nonclassical correlations improving the memristive features of our quantum device. The enhancement in the memristive properties could lead to improvements in neuromorphic algorithms if we replace the classical memristors by their quantum counterparts,  a good example is provided by ref. \cite{Spagnolo.2021} which employs quantum memristors for reservoir computing. Other application may involve analog simulation of differential equations \cite{Zidan.Nat}, or analog quantum neural networks composed of quantum memristors similar to the ones implemented with memristive crossbar latch architectures \cite{Miguel.Nat}.

Ordered behavior in nonlinear systems are useful in realizing neuromorphic architectures. This work advances the understanding of the interplay between memristivity and entanglement in open quantum systems. Furthermore, it lays the foundations for developing neuromorphic quantum computing and quantum neural networks with accessible superconducting technologies, on the way to reach quantum advantage with NISQ architectures.

\section{Acknowledgements}
The authors acknowledge support from Spanish PGC2018-095113-B-I00 (MCIU/AEI/FEDER), Basque Government IT986-16, projects QMiCS (820505) and OpenSuperQ (820363) of EU Flagship on Quantum Technologies, NSFC~(12075145), Shanghai~STCSM~(Grant 2019SHZDZX01-ZX04), and Program for Eastern Scholar. X.~C. acknowledges support from the Spanish Ram\'on y Cajal program (RYC-2017-22482).

%
\onecolumngrid
\appendix
\appendix

\section{Derivation of the circuit Hamiltonian}
\label{A}
The Lagrangian of the circuit shown in the Fig.~\ref{Fig:fig1} from main text is given by
\begin{eqnarray}\nonumber
\mathcal{L} = && \sum_{\ell=1,2}\bigg[\frac{C_{\Sigma,\ell}}{2}\dot{\varphi}_{\ell}^{2} + E_{J\ell}\cos\bigg(\frac{\varphi_{\ell}+\phi_{d\ell}(t)}{\varphi_{0}}\bigg) + E_{J\ell}\cos\bigg(\frac{\varphi_{\ell}+\phi_{d\ell}(t)+\phi_{s\ell}}{\varphi_{0}}\bigg)-\frac{\varphi_{\ell}^{2}}{2L_{\ell}}\bigg]\\
&& + \frac{C_c(\dot{\varphi}_{2}-\dot{\varphi}_{1})^{2}}{2} - \frac{(\varphi_{2}-\varphi_{1})^{2}}{2L_{c}},
\end{eqnarray}
where, $C_{\Sigma,1}=C_{J1}+C_{J2}$ and $C_{\Sigma,2}=C_{J3}+C_{J4}$ are the effective capacitance of the CA-SQUIDs, $E_{J\ell}$ is the Josephson energy of the $\ell$ junction, $\phi_{s\ell}$ is the static magnetic flux threading each SQUID, $L_{\ell}$ is the inductance of the $\ell$th quantum memristor, $\phi_{d\ell}(t)$ is the time dependent magnetic flux thought each quantum memristor, and $\varphi_0=\hbar/2e$ is the reduced flux-quanta. Finally, $C_c$ and $L_{c}$ are the coupling capacitance and inductance, respectively. Notice that for static magnetic flux at $\phi_{s\ell}/\varphi_{0}=\pi$ the Josephson energy contributions cancel each other, and the system Lagrangian simplifies to
\begin{eqnarray}
\mathcal{L}=\sum_{\ell=1,2}\bigg[\frac{(C_{\Sigma,\ell}+C_{c})}{2}\dot{\varphi}_{\ell}^{2} -\bigg(\frac{1}{2L_{\ell}}+\frac{1}{2L_{c}}\bigg)\varphi_{\ell}^{2}\bigg] - C_c\dot{\varphi}_{1}\dot{\varphi}_{2} + \frac{\varphi_{1}\varphi_{2}}{L_{c}}.
\end{eqnarray}
Now, the conjugate momenta $q_{\ell}=(\partial\mathcal{L}/\partial\dot{\varphi_{\ell}})$ are
\begin{eqnarray}
q_{1}&=&(C_{\Sigma,1}+C_{c})\dot{\varphi}_{1}-C_{c}\dot{\varphi}_{2} \, , \\
q_{2}&=&-C_{c}\dot{\varphi}_{1}+(C_{\Sigma,2}+C_{c})\dot{\varphi}_{2} \, .
\end{eqnarray}
We can write $\vec{q}=\hat{C}\vec{\dot{\varphi}}$, where $\vec{q}=(q_1,q_2)$, and $\vec{\varphi}=(\varphi_{1},\varphi_{2})$ and
\begin{eqnarray}
\hat{C}=\left(
\begin{array}{cc}
 C_{\Sigma,1} + C_c & -C_c \\
 -C_c & C_{\Sigma,2} + C_c  \\
\end{array}
\right) \, .
\end{eqnarray}
The Hamiltonian of the circuit is given by the transformation $\mathcal{H}=\vec{q}^{~~T}\vec{\dot{\varphi}}-\mathcal{L}=\vec{q}^{~~T}\hat{C}^{-1}\vec{q}-\mathcal{L}$, then
\begin{eqnarray}\nonumber
\mathcal{H}&=&\sum_{\ell=1,2}\bigg[\frac{\hat{C}^{-1}_{\ell,\ell}q_{\ell}^{2}}{2} +\frac{\hat{L}^{-1}_{\ell,\ell}\varphi_{\ell}^{2}}{2}\bigg] + \hat{C}^{-1}_{1,2}q_{1}q_{2} -\hat{L}^{-1}_{1,2}\varphi_{1}\varphi_{2} \, .
\end{eqnarray}
Here, $\hat{C}^{-1}_{j,k}$, and $\hat{L}^{-1}_{j,k}$ are the matrix element $(j,k)$ of the inverse of the capacitance and inductance matrix given by
\begin{eqnarray}
\hat{C}^{-1}=\frac{1}{C^{\star}}\left(
\begin{array}{cc}
 C_c+C_{\text{$\Sigma $2}} & C_c \\
 C_c & C_c+C_{\text{$\Sigma $1}} \\
\end{array}
\right), \qquad \hat{L}^{-1}=\left(
\begin{array}{cc}
 \frac{1}{L_c}+\frac{1}{L_1} & -\frac{1}{L_c} \\
 -\frac{1}{L_c} & \frac{1}{L_c}+\frac{1}{L_2} \\
\end{array}
\right),
\end{eqnarray}
with $C^{\star}=C_c \left(C_{\text{$\Sigma $1}}+C_{\text{$\Sigma $2}}\right)+C_{\text{$\Sigma $1}} C_{\text{$\Sigma $2}}$. To quantize the circuit Hamiltonian, we promote the variables to operators. It means, $q_{\ell}\rightarrow\hat{q}_{\ell}=-2e\hat{n}_{\ell}$, and $\varphi_\ell/\varphi_0\rightarrow\hat{\phi}_{\ell}$, with $[\hat{\phi}_{\ell},\hat{n}_{\ell'}]=i\delta_{\ell,\ell'}$. Therefore, the quantum Hamiltonian reads
\begin{eqnarray}
\label{Hamiltonian}
\mathcal{H}&=&\sum_{\ell=1,2}\bigg[E_{C_{\ell,\ell}}\hat{n}_{\ell}^{2} +\frac{E_{L_{\ell,\ell}}}{2}\hat{\phi}_{\ell}^{2}\bigg] + E_{C_{1,2}}\hat{n}_{1}\hat{n}_{2} -E_{L_{1,2}}\hat{\phi}_{1}\hat{\phi}_{2},
\label{simpHamiltonian}
\end{eqnarray}
where $E_{C}=2e^2\hat{C}^{-1}$ is the charge energy and $E_{L}=\varphi_{0}^{2}\hat{L}^{-1}$ is the inductive energy. For simplicity, we can redefine $\hat{q}_{\ell}$ and $\hat{\phi}_{\ell}$ in terms of creation and annihilation operators as

\begin{eqnarray}
\hat{n}_{\ell}&=&\frac{i}{4g_{\ell}}(a_{\ell}^{\dag}-a_{\ell}),\\
\hat{\phi}_{\ell}&=&2g_{\ell}(a_{\ell}^{\dag}+a_{\ell}),
\end{eqnarray}
where $g_{\ell}=(E_{C_{\ell,\ell}}/32E_{L_{\ell,\ell}})^{1/4}$ is proportional to the zero-point fluctuation of the phase operator. Finally, the quantum Hamiltonian reads
\begin{eqnarray}
\label{Hamiltonian}
\label{Quantum_H}
\hat{\mathcal{H}}=\sum_{\ell=1,2}^1\hbar\omega_\ell \hat{a}_\ell^{\dagger}\hat{a}_\ell - \sqrt{\omega_{1}\omega_{2}}(\alpha-\beta)(\hat{a}_1^{\dagger}\hat{a}_2 + \hat{a}_1\hat{a}_2^{\dagger}),
\end{eqnarray}
where, $\omega_{\ell}=\sqrt{2E_{C_{\ell,\ell}}E_{L_{\ell,\ell}}}/\hbar$ is the frequency of the $\ell$th QMs. Moreover, $\alpha=E_{L_{1,2}}/\sqrt{E_{L1}E_{L2}}$, and $\beta=E_{C_{1,2}}/\sqrt{E_{C1}E_{C2}}$ stand for the inductive and charge energy ratios.

\section{Memristive equations}
\label{B}
We derive the equation of motion for the mean value of the observables $\hat{n}_{\ell}$, and $\hat{\phi}_{\ell}$, which are related to the current flowing and the voltage across the memristor. We will consider the dynamics in the Schr\"odinger picture, where it is governed by the following master equation
\begin{eqnarray}
\label{Mst_eq_1}
\dot{\hat{\rho}}(t) &=& -\frac{i}{\hbar}\big[\mathcal{H},\hat{\rho}\big] + \sum_{\ell=1,2}\frac{\Gamma_{\ell}(t)}{2}\bigg[\mathcal{L}_{\ell}\hat{\rho}\mathcal{L}^{\dag}-\frac{1}{2}\{\mathcal{L}^{\dag}\mathcal{L},\hat{\rho}\}\bigg] \, .
\end{eqnarray}
Here, $\hat{\rho}(t)$ is the density matrix describing the system and $\dot{\hat{\rho}}(t)=d\hat{\rho}(t)/dt$ represents the derivative with respect to time. Moreover, $\mathcal{H}$ is the system Hamiltonian in Eq. (\ref{Hamiltonian}), while $\mathcal{L}_{\ell}=\sqrt{\Gamma_{\ell}(t)}a_{\ell}$ is the collapse operator describing the quasiparticle tunneling of the $\ell$th QMs at the rate $\Gamma_{\ell}(t) = \lvert \bra{0}\sin(\phi_{\ell}/2)\ket{1}\lvert^{2}S_{\rm{qp}}(\omega_{\ell})$. \\ We are interested in the time evolution of the expectation value $\langle\hat{\mathcal{O}}(t)\rangle = {\rm{Tr}}[\hat{\mathcal{O}}\hat{\rho}(t)]$ with $\hat{\mathcal{O}} = \{\hat{n}_{\ell}, \hat{\varphi}_{\ell} \}$. Using Eq.~(\ref{Mst_eq_1}), we obtain the equation of motion for $\hat{\mathcal{O}}$ as 
 \begin{eqnarray}
 \label{Chap3_Sec3_Eq3}
\frac{d\langle\hat{\mathcal{O}}(t)\rangle}{dt}= -\frac{i}{\hbar}{\rm{Tr}}\bigg[[\mathcal{H},\mathcal{O}]\hat{\rho}(t)\bigg] +{\rm{Tr}}\bigg[ \tilde{\mathcal{D}}[\mathcal{O}]\hat{\rho}(t)\bigg],
 \end{eqnarray}
where $\tilde{\mathcal{D}}[\mathcal{O}]=\Gamma(t)(a^{\dag}\mathcal{O} a - \{a^{\dag}a,\mathcal{O}\}/2)$ corresponds to the Lindbladian for the operator $\mathcal{O}$. For $\hat{n}_{k}$, we obtain 

\begin{subequations}
\begin{eqnarray}
\label{Chap3_Sec3_Eq4b}
\frac{d\langle\hat{n}_{k}\rangle}{dt} &=&E_{L_{k,k}}\langle\hat{\phi}_{k}\rangle - E_{L_{1,2}}(\langle\hat{\phi}_{1}\rangle\delta_{2,k} - \langle\hat{\phi}_{2}\rangle\delta_{1,k}) - {{\rm{Tr}}}\bigg[\tilde{\mathcal{D}}[\hat{n}]\hat{\rho}(t)\bigg]
\end{eqnarray}
\end{subequations}
and
\begin{eqnarray}
\tilde{\mathcal{D}}[\hat{n}_{k}]=\frac{4i\Gamma(t)}{g_{0}}\bigg[-\frac{a_{k}^{\dag}}{2} +\frac{a_{k}}{2}\bigg]=-\frac{\Gamma(t)}{2}\hat{n}_{k} \, .
\end{eqnarray}
Thus, the equation of motion reads
\begin{eqnarray}
\frac{d\langle\hat{n}_{k}\rangle}{dt} &=&E_{L_{k,k}}\langle\hat{\phi}_{k}\rangle - E_{L_{1,2}}(\langle\hat{\phi}_{1}\rangle\delta_{2,k} - \langle\hat{\phi}_{2}\rangle\delta_{1,k}) -\frac{\Gamma(t)}{2}\langle\hat{n}_{k}\rangle.
\end{eqnarray}
Analogously for $\hat{\phi}_{k}$, we obtain
\begin{eqnarray}
\frac{d\langle\hat{\phi}_{k}\rangle}{dt} &=&-2E_{C_{k,k}}\langle\hat{n}_{k}\rangle - E_{C_{1,2}}(\langle\hat{n}_{1}\rangle\delta_{2,k} + \langle\hat{n}_{2}\rangle\delta_{1,k}) -\frac{\Gamma(t)}{2}\langle\hat{\phi}_{k}\rangle.
\end{eqnarray}

\section{Inductively coupled quantum memristors}
\label{C}
\begin{figure}[!t]
\centering
\includegraphics[width=0.8\linewidth]{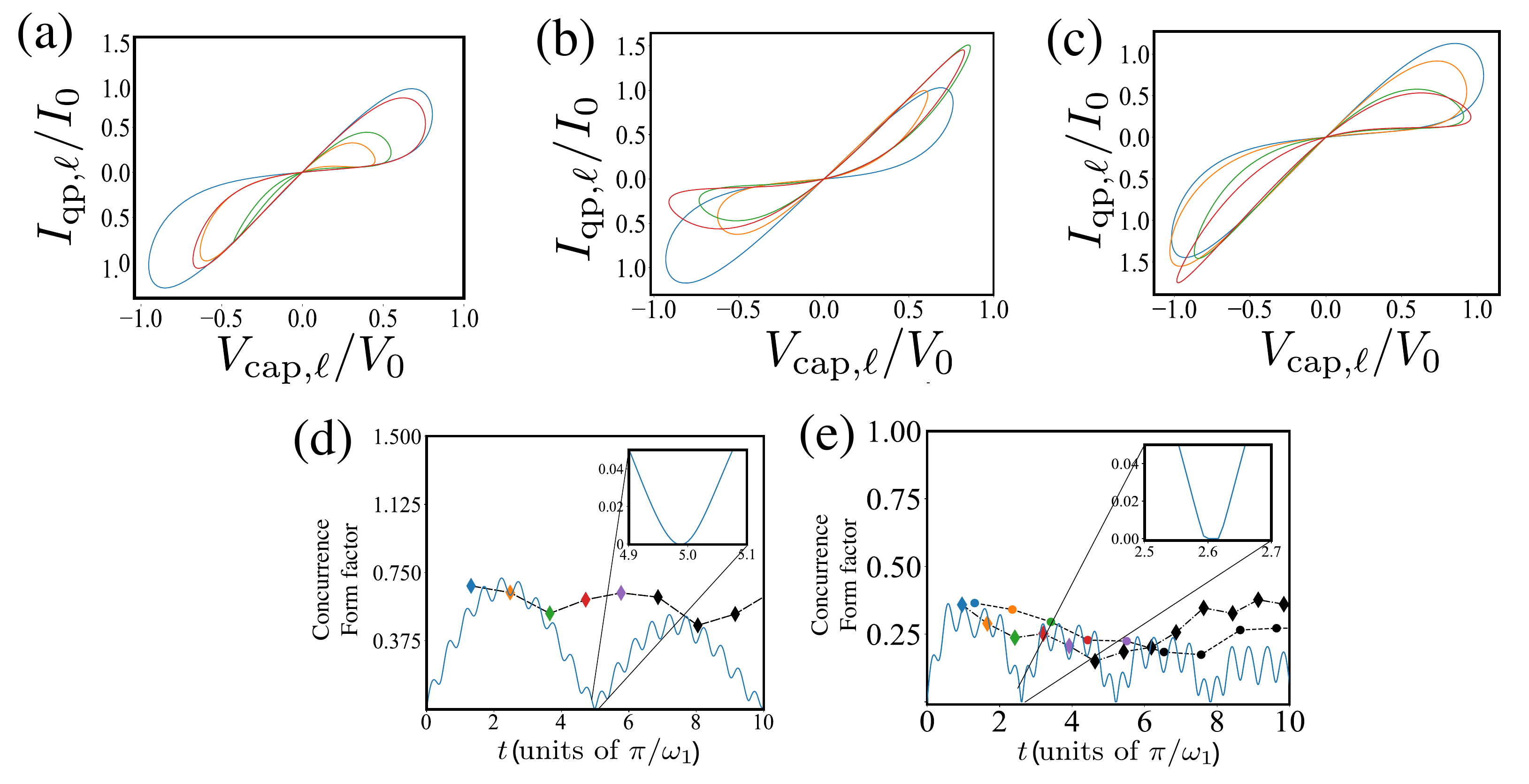}
\caption{Inductively coupled QMs. Entanglement and memristive dynamics for the case of identical memristors (a), (d) and non-identical memristors (b), (c), (e). Memristive dynamics are shown for five oscillations using five different colors. The corresponding form factor (normalized with respect to the concurrence) is plotted using the same color combination of points for each oscillation along with the concurrence. The parameter choice and initialization is the same as in Fig.~\ref{Fig:fig3} of the main text.}
\label{fig6}
\end{figure}

In this section, we discuss the results of the inductive interaction between the QMs. We omit the dynamic response and the performance since it is included in the main text for the capacitively coupled QMs. Here, we show that the inverse relation between entanglement and memristive dynamics is also satisfied for the inductive interaction along with the observation of ESD and ESB. 

Fig. \ref{fig6} shows the entanglement and memristive dynamics for the QMs coupled via inductive interaction. The form factor is shown along with the concurrence with the same color points as their corresponding pinched hysteresis loop. We observe that the coupling leads to the periodic shrink and expansion of the hysteresis curve along with the periodic rise and decay of the concurrence. To reflect upon this fact, we can compare the timescales of the form factor and the concurrence for identical QMs, Fig. \ref{fig6} (d), where we can find that the loss of entanglement is accompanied by the increase in the form factor. It hints that that the entanglement and memristive dynamics may generally be inversely related to each other. Furthermore, the interaction leads to the competition between the input voltage update time ($\tau_{1}=2\pi/\omega_{\ell}$) and the enveloping time ($\tau_{2}=2\pi/g$) of the QM inducing ESD (shown in the insets) and ESB which can be observed for both the cases. On the other hand, similar to the capacitive coupling, when we detune the QMs, Fig. \ref{fig6} (e), the distribution of information exchange between them are uneven due to which the shrink/expansion of the hysteresis and the corresponding rise/decay of the quantum correlations do not happen in the same timescale. 

In conclusion, despite the minor differences in the plots of the inductive and capacitive coupling, which are mainly due to the very nature of the circuit elements, the interaction produces correlated inputs during the system evolution leading to periodic shrink/expansion of the hysteresis and the rise/decay of the entanglement. The inverse behavior of the entanglement and memristive dynamics can be generally observed when the detuning is low.

\begin{figure}[!th]
\centering
\includegraphics[width=0.8\linewidth]{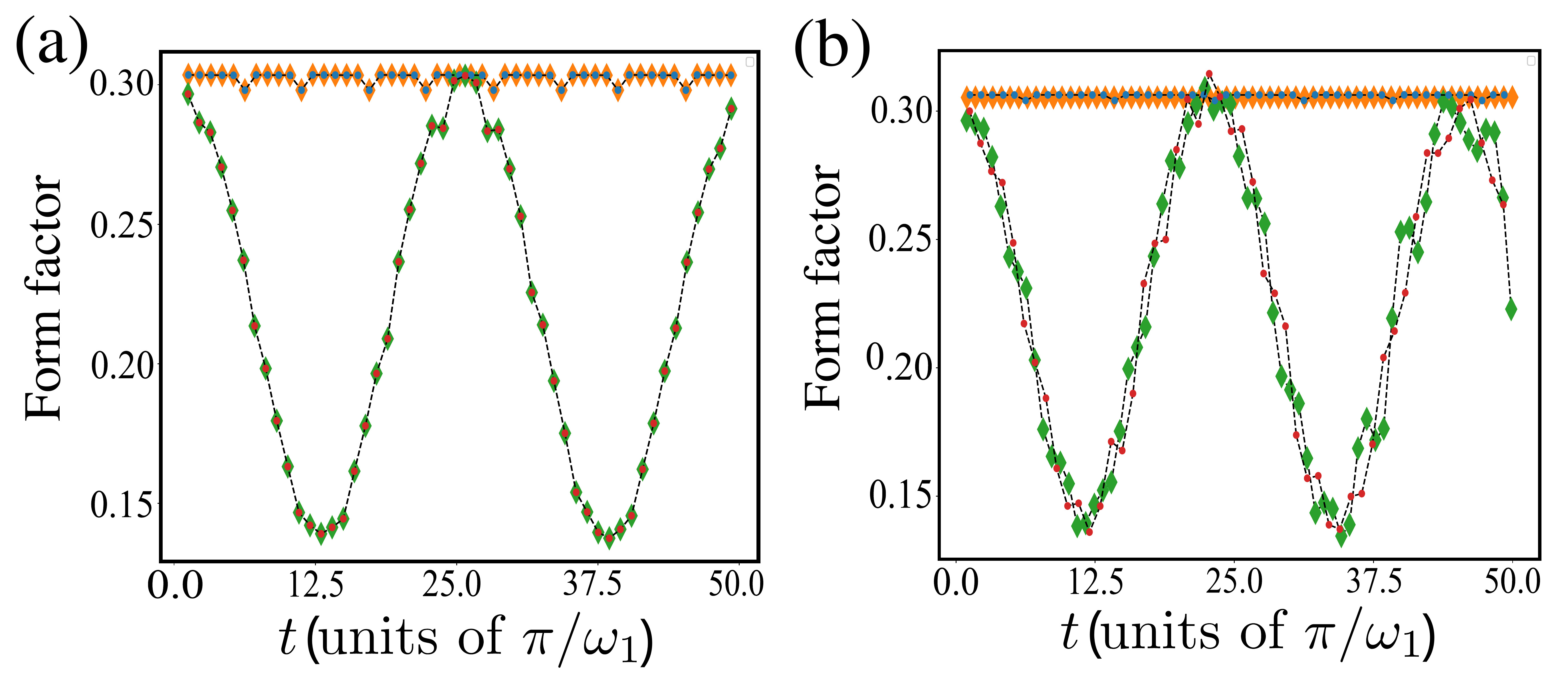}
\caption{QMs coupled via an inductor and a capacitor simultaneously. Form factor (area/perimeter$^2$) of the hysteresis curves as a function of the period for (a) identical and (b) non-identical QMs. QM$_1$ and QM$_2$ are shown using diamonds and dots, respectively. The parameter choice and initialization is the same as in Fig.~\ref{Fig:fig3} of the main text.}
\label{Fig.7}
\end{figure}

\begin{figure}[!th]
\centering
\includegraphics[width=0.8\linewidth]{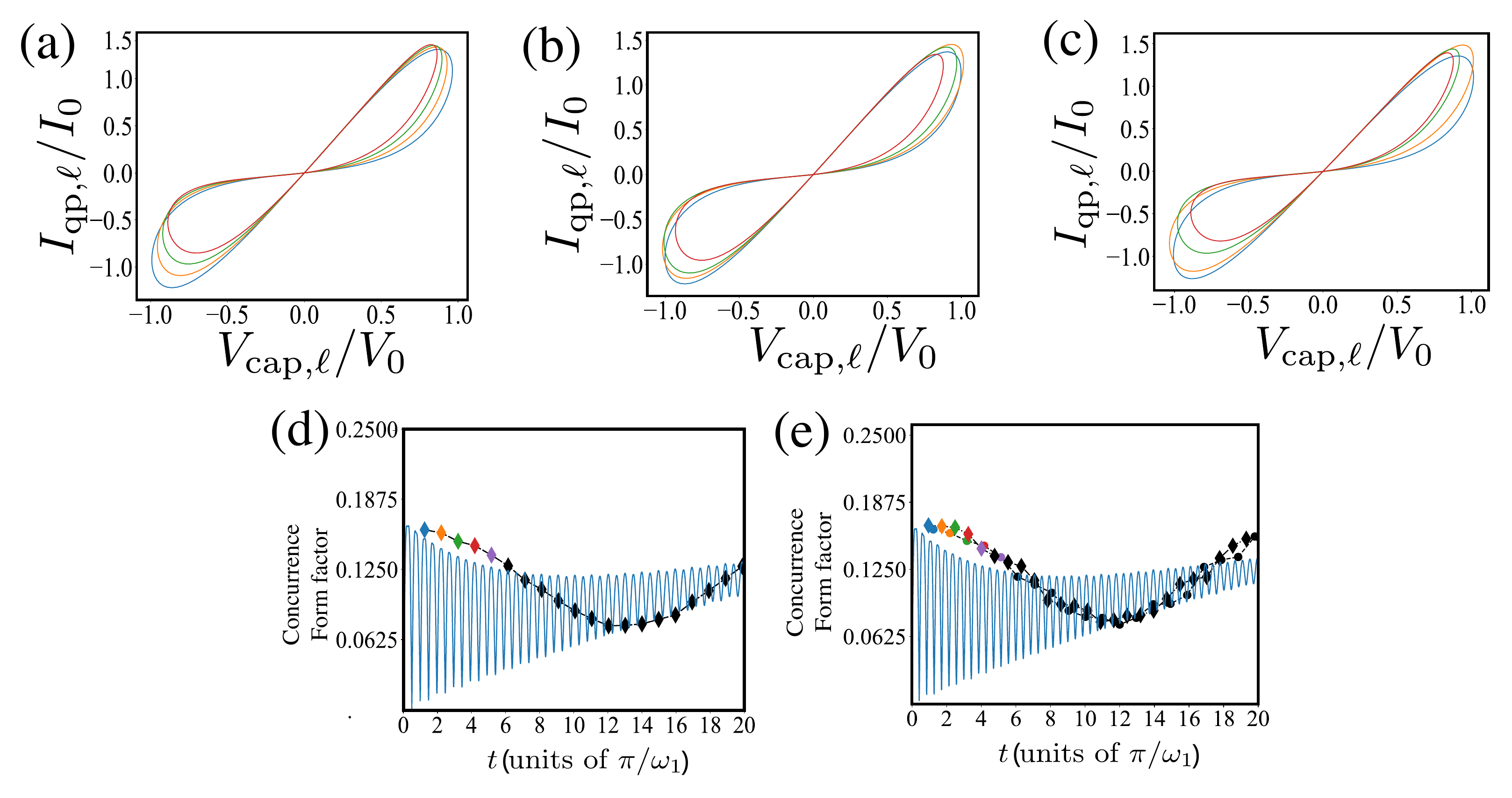}
\caption{QMs coupled via an inductor and a capacitor simultaneously. Entanglement and memristive dynamics for the cases of identical QMs (a), (d) and non-identical QMs (b), (c), (e) . Memristive dynamics are shown for four oscillations in the timescale of QM$_1$ using different colors. The corresponding form factor is plotted using the same color combination of points for each oscillation along with the concurrence. The parameter choice and initialization is the same as in Fig.~\ref{Fig:fig3} of the main text.}
\label{Fig.8}
\end{figure}

\begin{figure*}[!th]
\centering
\setlength{\belowcaptionskip}{-10pt} 
\includegraphics[width=1\linewidth]{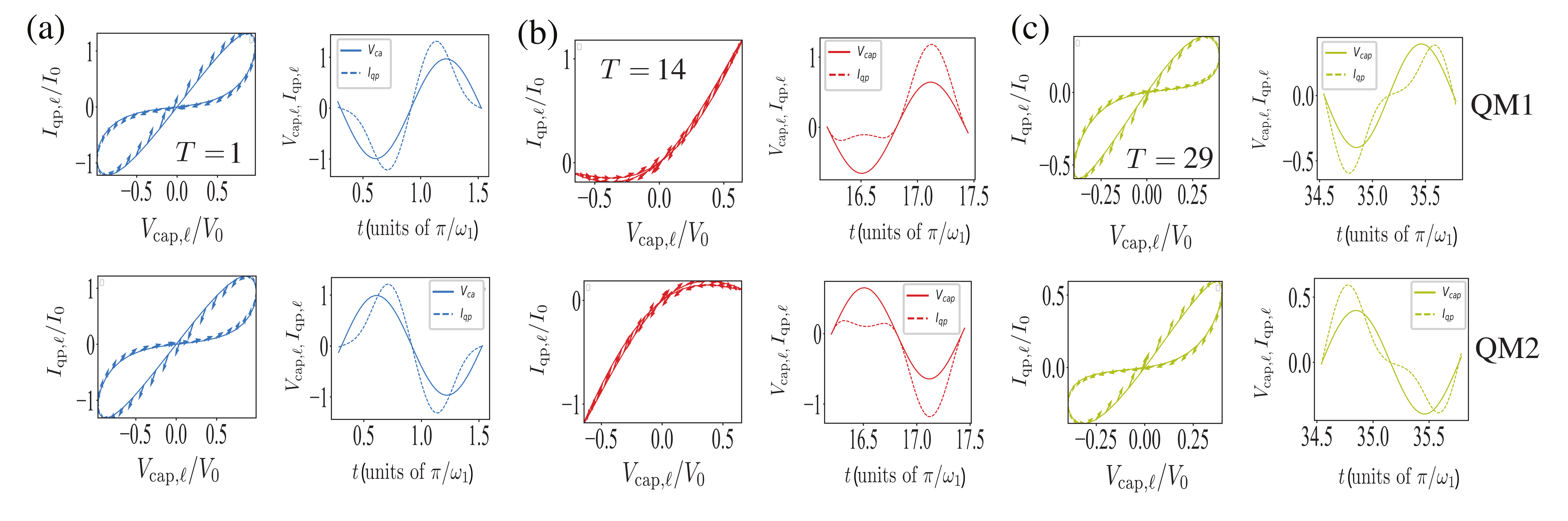}
\caption{QMs coupled via an inductor and a capacitor simultaneously. Direction of the hysteresis curve(left) and corresponding time dependence of current and voltage(right) at the (a)~first oscillation, (b)~14th oscillation and (c)~29th oscillation for the case of identical memristors. The first(second) row corresponds to $\textrm{QM}_{1(2)}$. We can see that the relative phase between voltage and current changes in time. In the 14th oscillation (b), this phase has reversed, which changes the direction of the hysteresis curve and the relative phase is inverted in the 29th oscillation (c). We have used the same initial state and system parameters as in the Fig. \ref{Fig:fig3}.}
\label{Fig.9}
\end{figure*}

\section{Quantum memristors coupled via an inductor and a capacitor simultaneously}
\label{D}

Apart from using either an inductor or a capacitor, the coupling characteristics can be investigated when both the circuit elements are simultaneously used to couple the QMs. Thus, the system is fully described by the simplified and the quantized hamiltonian given in the Eqn. (\ref{simpHamiltonian}) and Eqn. (\ref{Quantum_H}) with both the inductive and capacitive coupling contributions. For this system we study the performance and the entanglement/memristive dynamics in order to compare it with the other coupling schemes. 

The performance, quantified by the form factor ($\mathcal{F}=4\pi\frac{A}{P^2}$) is shown in the Fig. \ref{Fig.7}, for identical and nonidentical cases. As we can see, the coupling leads to correlated inputs inducing rise and decay of the form factor. In comparison to the inductive/capacitive coupling, the oscillation frequency is low. Here the coupling doesn't lead to an enhanced performance of the QMs. Moreover, the nonidentical ones do not show a significant change in their behavior as compared to the identical case. This observation can be understood by investigating the interplay of the entanglement and memristive dynamics as shown in the Fig. \ref{Fig.7}. Similar to the other cases, the periodic shrink/expansion of the hysteresis curves and the rise/decay of the concurrence is observed. We see that there's no observation of the ESD/ESB phenomena. The entanglement in this case is very low due to which the maximal/minimal of concurrence/form factor never coincide and their oscillation frequencies are not comparable. Therefore, here we observe that the coupling leads to correlated inputs but there's no observation of ordered characteristics as in the coupling via a single inductor/capacitor. Nonetheless, the  direction of the hysteresis is reversed during the time evolution of the system. As can be seen in the Fig. \ref{Fig.9} (a), the two QMs initialized out of phase have their relative phases inverted during the time evolution of the system Fig. \ref{Fig.9} (c). This inversion occurs when the entanglement tends towards a maximum value after undergoing net decay (compare with Fig. \ref{Fig.8}).

\appendix
\appendix

\end{document}